\documentclass[11pt]{article}
\usepackage{moriond,epsfig,floatflt}

\def\be{\begin{equation}}
\def\ee{\end{equation}}
\def\bea{\begin{eqnarray}}
\def\eea{\end{eqnarray}}

\def\gev{\ensuremath{\mathrm{GeV}}}
\def\ipb{\ensuremath{\mathrm{pb}^{-1}}}
\begin{document}
\vspace*{4cm}
\title{FINAL STATE INTERACTIONS IN HADRONIC WW DECAY AT LEP}

\author{MARK DIERCKXSENS}

\address{NIKHEF, PO Box 41882, 1009 DB Amsterdam, The Netherlands}

\maketitle\abstracts{An overview is given of the study of
  final state interactions in hadronically decaying W pairs produced
  in $e^+ e^-$-collisions as it is performed by the four LEP
  experiments. Bose-Einstein correlations are investigated  
  by comparing like- with unlike-signed pairs of pions and/or using
  the mixed event analysis technique. Colour reconnection is
  examined with a method that compares the particle flow
  distributions in inter-jet regions.} 

\section{Introduction}\label{sec:intro}

The second phase of the LEP programme was devoted to run at
centre-of-mass energies above the W pair threshold. When such a pair
is produced, the W bosons will immediately decay due to their very short
lifetime ($\tau_W \sim 0.1~\mathrm{fm/c} $). 
In $46\%$ of the cases this will give rise to a final state with four
quarks, the so-called hadronic channel. Considering the full integrated
luminosity delivered by the LEP machine, each experiment has collected
roughly 5000 such events.
The produced quarks will hadronise after typical scales of about
1 fm, resulting in jets of particles measured with the detector. 
Due to the large space-time overlap, correlations between the
particles coming from different W's might occur. 
The possible effects studied at LEP are
Bose-Einstein correlations (BEC) between identical bosons in the jets,
and colour reconnection (CR) between the quarks themselves.   

When an exchange of particles or energy and momenta between
the two W bosons is inadequately described by the Monte Carlo
programs, this will cause a bias in the measurement of the W mass.
The statistical uncertainty on the W mass is
$30~\mathrm{GeV/c^2}$ when combining the results from all the LEP
experiments using the hadronic channel only~\cite{lepew}.
Various models exist to describe the Final State Interactions (FSI)
and a number of them predict biases several times larger than this
statistical error. Besides the measurement on itself, these results
are therefore also important for the determination of the W mass.

\section{Bose-Einstein Correlations}\label{sec:be}

Since the bosons produced in the hadronisation process obey
Bose-Einstein statistics, an enhancement in the number of  
identical bosons created close in phase space is expected due to
amplitude symmetrisation. 
This effect has been observed in e.g.~$Z \rightarrow q\bar{q}$ at
LEP1, thus between particles coming from the same boson. Now that
boson pair production is possible at LEP2 energies, the question which
arises is whether BEC also exist between particles coming from
different W's (inter-W BEC). 
The energy ranges and luminosities used by the four LEP experiments
for this measurement~\cite{lepbe} can be found in table
\ref{tab:lumi}. 

\begin{table}[t]
  \caption{The energy ranges and corresponding integrated luminosities
    used by the four LEP experiments for the studies of Bose-Einstein
    correlations and colour reconnection.}
  \label{tab:lumi}
  \vspace{0.4cm}
  \begin{center}  
    \begin{tabular}{|c|cc|cc|}
      \hline
      Experiment & \multicolumn{2}{|c|}{BEC} & \multicolumn{2}{|c|}{CR} \\
      & \ $\sqrt{s}~[\gev]$ \ & \ $\mathcal{L}~[\ipb]$ \ &
      $\sqrt{s}~[\gev]$ & $\mathcal{L}~[\ipb]$ \\
      \hline
      ALEPH  & $183 - 209$ & 683 & $189 - 209$ & 628 \\ 
      DELPHI & $189 - 209$ & 531 & $183 - 209$ & 601 \\
      L3     & $189 - 209$ & 627 & $189 - 209$ & 627 \\
      OPAL   & $172 - 189$ & 250 & $189$       & 183 \\
      \hline
    \end{tabular}
  \end{center}
\end{table}

BEC are typically studied by looking at the two-particle density in
momentum space
\begin{equation}
  \rho(Q) = \frac{1}{N_{ev}}\frac{\mathrm{d}n_{pairs}}{\mathrm{d}Q},
  \label{eq:rho}
\end{equation}
where $n_{pairs}$ is the number of pairs in $N_{ev}$ selected events
and Q is the 4-momentum difference $\sqrt{-(p_1-p_2)^2}$, with $p_1$
and $p_2$ the momenta of the two particles. The presence
of BEC will give rise to an enhancement of like-signed particle pairs
at low Q. The two-particle density is often parametrised as $\rho \simeq
(1+ \Lambda \exp(-R^2 Q^2))$ and interpreted as a Gaussian source of
radius $R$ and strength $\Lambda$. 

The $BE_{32}$ algorithm from the LUBOEI model~\cite{luboei} is used to
generate Monte Carlo samples containing 
BEC inside and between the W bosons. This model just reshuffles the
momenta of identical bosons and is tuned to the hadronic Z decay data
to describe BEC in these events.

\subsection{Unlike-sign Pair Analysis}

In this method, the ratio between the number of like-signed pairs and
unlike-signed pairs obtained from data is compared with a Monte Carlo sample
generated without inter-W BEC and studied as a
function of Q.  

\begin{figure}[tbhp] 
 \centerline{\mbox{\epsfig{file=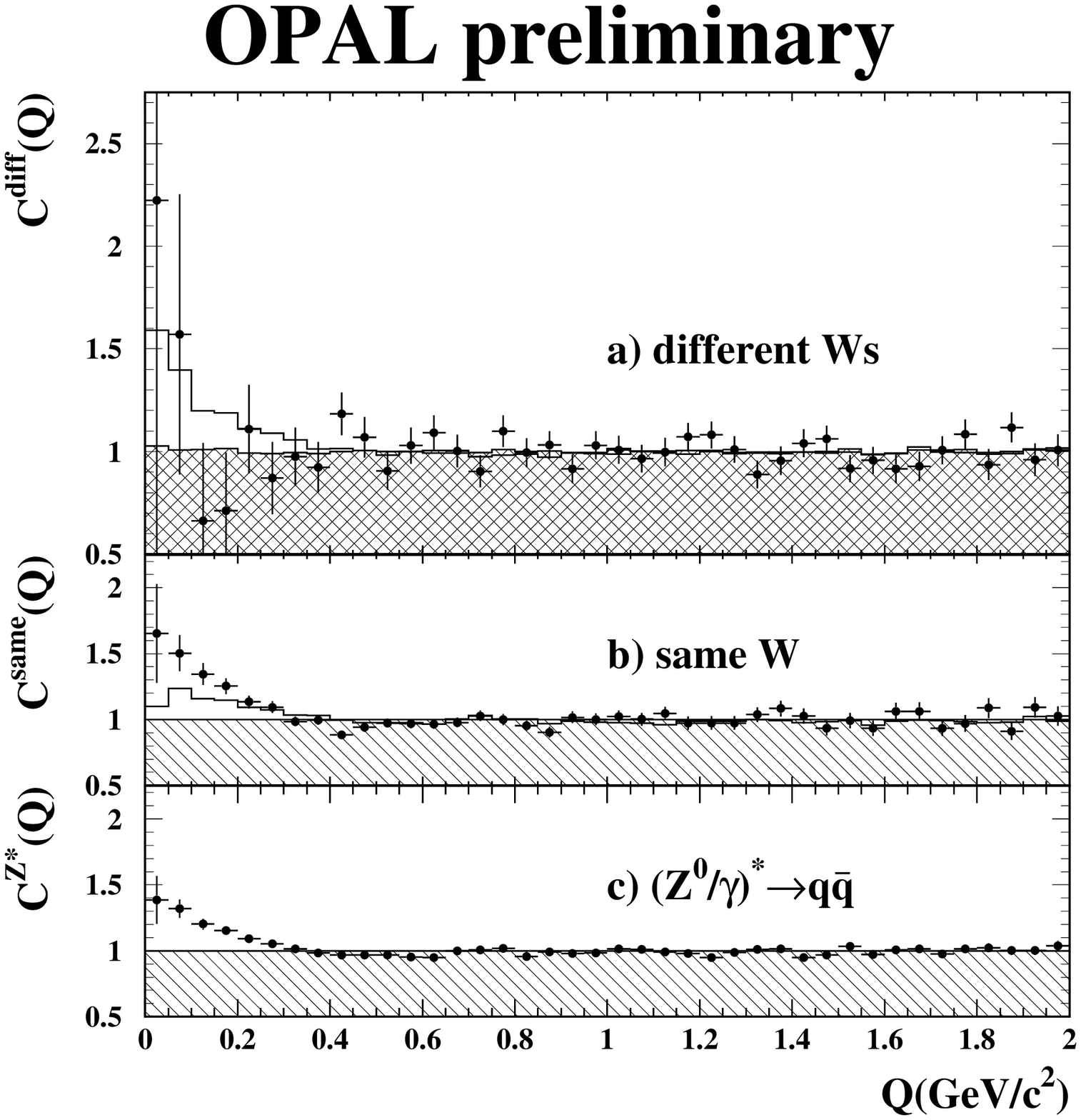,width=0.4\textwidth}
     \epsfig{file=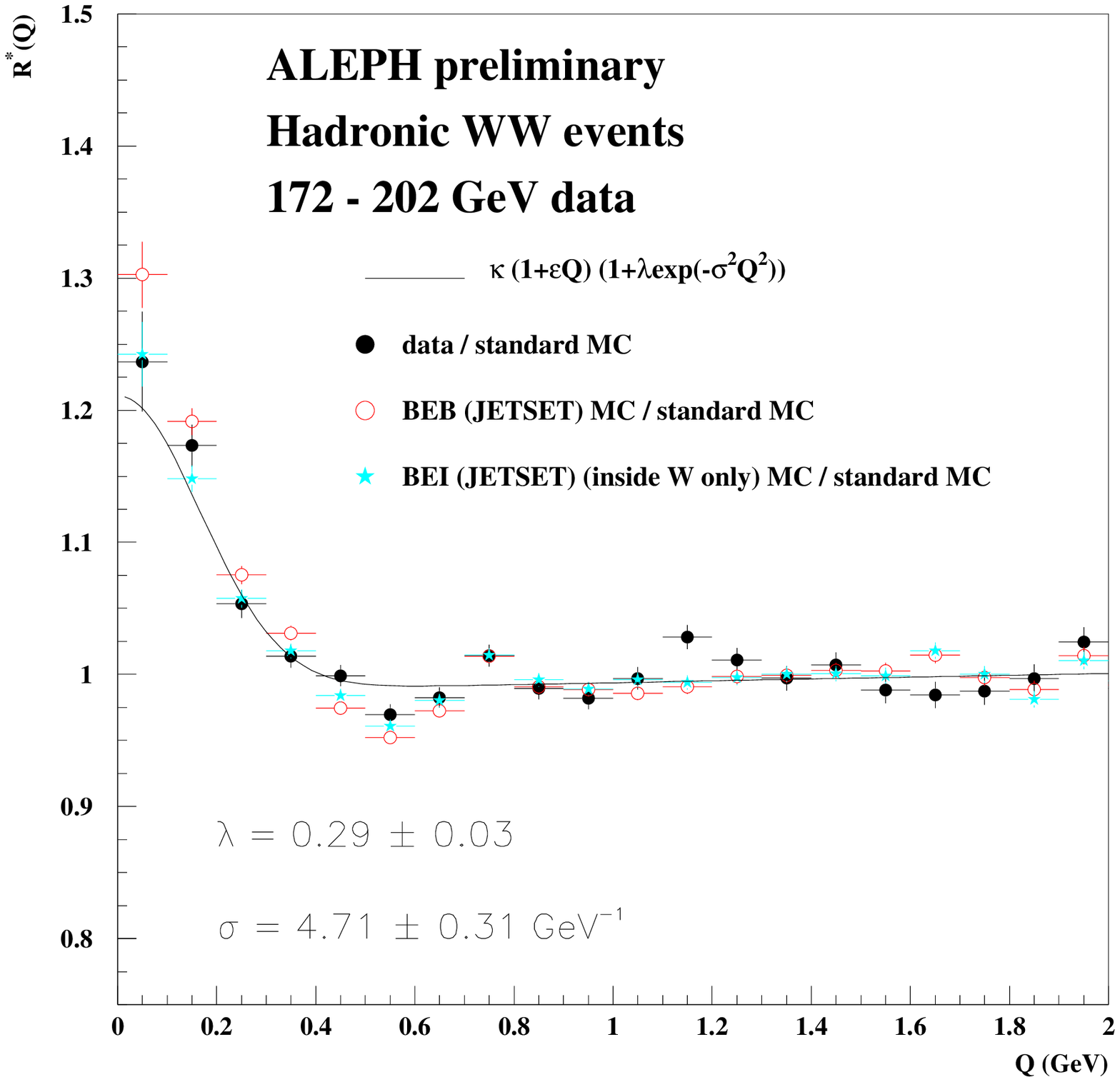,width=0.4\textwidth}}}
 \caption{The double ratio distributions as a function of $Q$ from
     OPAL (left), 
     deconvoluted to different contributions, and ALEPH (right), where
     the fit to the data is indicated as a solid line.} 
 \label{fig:pairs}
\end{figure}

OPAL extracts these values for three event classes: hadronic and
semi-leptonic WW decays and non-radiative $Z^* \rightarrow 
q\bar{q}$ events. The double ratios are then deconvoluted to obtain
the contributions of pions coming from different W's, same W and
hadronic Z decays. The result is shown in figure \ref{fig:pairs}. A
fit to these distributions results in a strength parameter
$\Lambda = 0.69 \pm 0.12 \pm 0.06$ for pairs coming from the
same W, again proving the existence of BEC inside the W. The fit to
the distribution for pions from different W's gives $\Lambda = 0.05 \pm
0.67 \pm 0.35$, being compatible with both inter-W and no inter-W BEC
due to the small statistical sample used.

The double ratio from the ALEPH data up to 202 GeV shown in the right
plot of figure \ref{fig:pairs}, is fitted with a similar function as
described above and the integral over Q of this function is
calculated. The same procedure is followed for a MC sample using the
JETSET model with inter-W BEC. When compared to the data, this model
is disfavoured at the level of $2.2\sigma$.

\subsection{Mixed Event Analysis}

Inter-W BEC can be accessed directly from data, without relying on a
certain model, by comparing the hadronic decays
with events constructed from the hadronic part of two different
semi-leptonic events~\cite{mix}. This analysis has been preformed by
ALEPH, DELPHI and L3. 

If the two W bosons in a hadronic event decay
independently, the two-particle density can be written as 
\begin{equation}
  \rho^{WW \rightarrow 4q} = 2 \rho^{W \rightarrow 2q} + 2
  \rho^{WW}_{mix},
  \label{eq:rhoww}
\end{equation}
where the first term on the right hand side is taken from
semi-leptonic events, and the second term is obtained by mixing the
hadronic part of two different semi-leptonic events. Then the sensitive
distributions are the difference or the ratio between left and
right hand side:
\begin{equation}
  \Delta \rho = \rho^{WW \rightarrow 4q} - 2 \rho^{W \rightarrow 2q} -
  2 \rho^{WW}_{mix}, \qquad D = \frac {\rho^{WW \rightarrow 4q}} { 2
  \rho^{W \rightarrow 2q} + 2 \rho^{WW}_{mix}}. 
  \label{eq:drd}
\end{equation}
When BEC between particles coming from different
W bosons do not exist, $\Delta \rho = 0$ and $D = 1$.

\begin{figure}[bthp]
\centerline{\epsfig{file=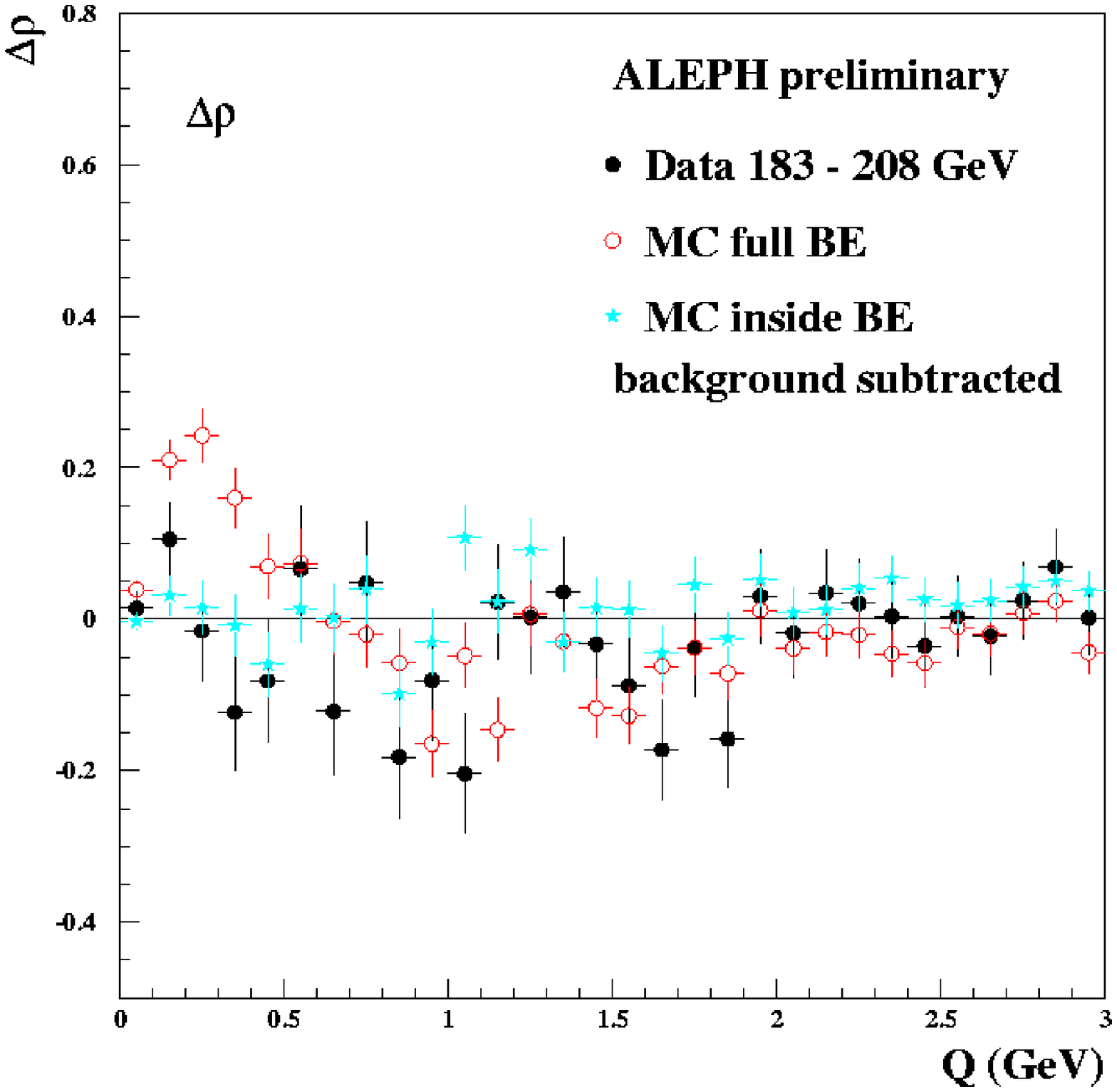,width=0.38\textwidth} 
  \epsfig{file=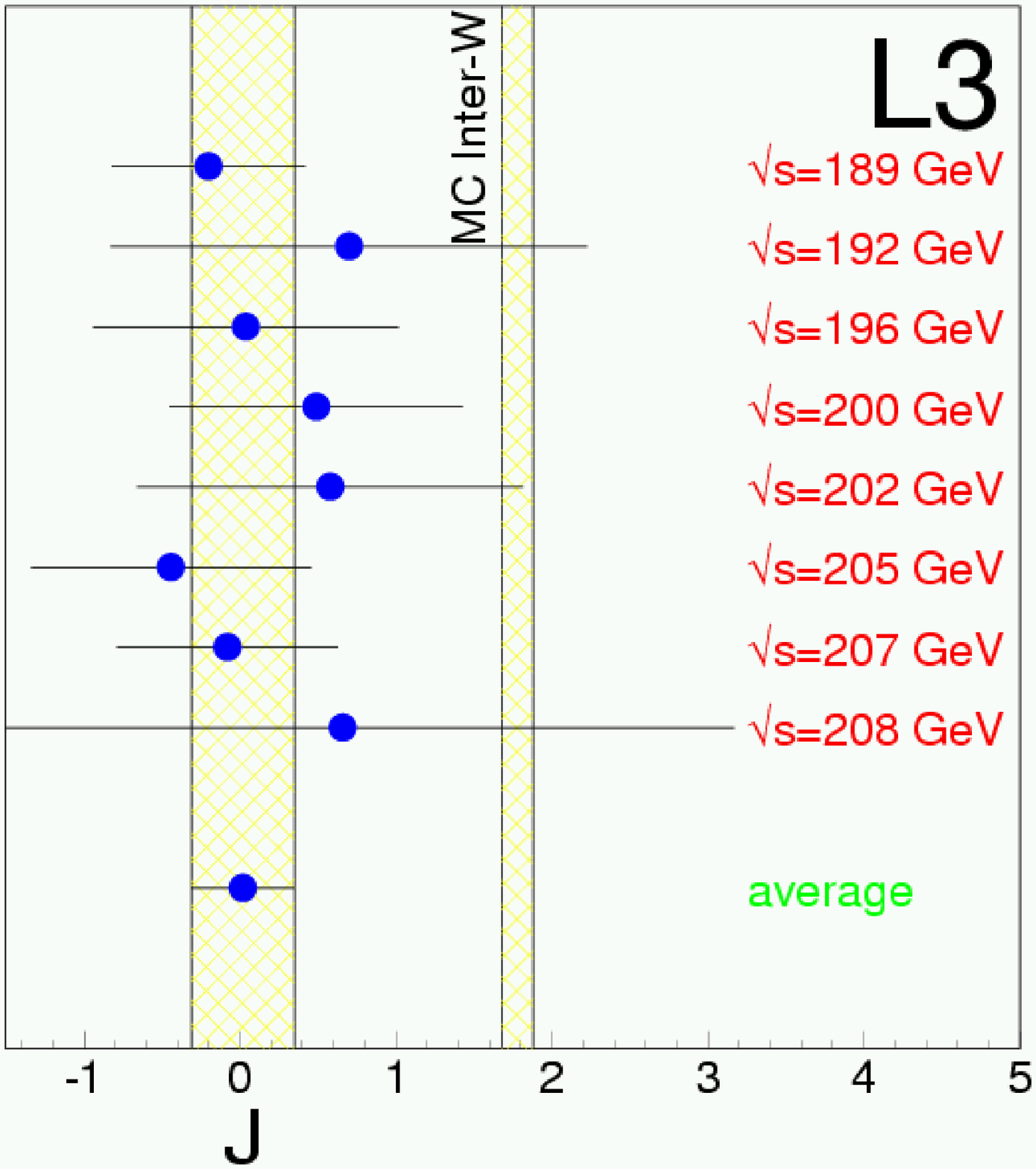,width=0.36\textwidth}}
\caption{The $\Delta \rho$ distribution as a function of $Q$ from
  ALEPH (left) and the 
  integral $J \equiv \int_{0}^{Q_{max}} \Delta \rho (Q) 
  \mathrm{d}Q$ at different centre-of-mass energies from L3 (right).}  
\label{fig:rdist}
\end{figure}

Figure \ref{fig:rdist} shows the $\Delta \rho$ distribution for the
ALEPH data, which is in good agreement with the prediction from a MC
containing only intra-W BEC. The same conclusions can be drawn when
looking at these distributions from DELPHI and L3. 
Another way to look
at the data is to calculate the integral $J 
\equiv \int_{0}^{Q_{max}} \Delta \rho (Q) \mathrm{d}Q$. As can be seen
in the right plot of figure \ref{fig:rdist}, the value obtained from a
MC with inter-W BEC 
clearly does not agree with the data. 

\begin{floatingfigure}[r]{7.5cm}
\centerline{
\epsfig{file=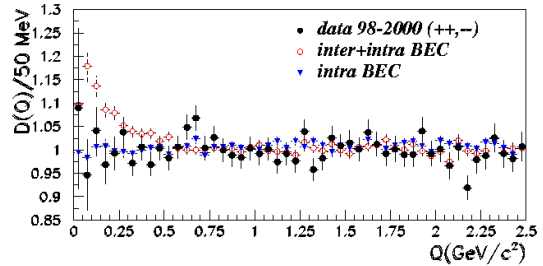,width=0.45\textwidth}
 }
\caption{The $D$ distribution as a function of $Q$ from the DELPHI
  experiment}   
\label{fig:ddist}
\end{floatingfigure}

Comparing the
strength parameter $\Lambda$ from the fit to the
distribution of the ratio D from the DELPHI experiment
(fig.~\ref{fig:ddist}), the MC containing inter-W BEC is excluded on
the level of $3.2 \sigma$. L3 looks at the ratio between the D
distribution from data and this distribution from a MC with only
intra-W BEC in order to eliminate some possible systematic
errors. Comparing the strength parameter from the fit to this double 
ratio using the data with the one obtained when a MC with inter-W BEC
is used, the latter one can be excluded by $4.7$ standard deviations.

\section{Colour Reconnection}\label{sec:cr}

In the string picture, the hadronisation particles come
from the decay of the colour string stretched between the quarks
coming from the same W.
When the colour flow pattern is modified, referred to as colour
reconnection, strings might span 
between two quarks from different W bosons and sebsequently decaying
particles cannot be uniquely assigned to either W. 

Different phonomenological models exist to describe CR. In the model
of Sj\"ostrand-Khoze~\cite{sk}, as it is implemented in PYTHIA, CR
might occur when strings overlap. In the type I model (SKI), the
strings are associated to colour flux tubes and the reconnection
probability, $P_{reco}$, is related to the volume of the overlap,
$f(\sqrt{s})$, by the following relation:  
\begin{equation}
  \label{eq:ski}
  P_{reco} = 1 - e^{-f(\sqrt{s}) \cdot k_I}, 
\end{equation}
where $k_I$ is a free parameter that can be varied to obtain samples
with different reconnection probabilities.
Also the predictions from other models like SKII, ARIADNE, HERWIG and
Rathsman are tested against the LEP data. 

\subsection{Particle Flow Method}

This method compares the particle rates between jets from same and 
different W bosons, since CR will result in a depletion or 
enhancement in inter-jet regions. This approach, pioneered by
L3~\cite{duch}, is much more sensitive to CR effects than previous 
methods and is now adopted by the other experiments. The data samples
used by the different LEP experiments for this 
analysis~\cite{lepcr} are indicated in table \ref{tab:lumi}.

In order to study the particle flow, the momenta of the particles are
projected onto the plane spanned by the most energetic jet and the jet
most likely coming from the same W. 
Since the W decay planes are not planar, the momentum of a particle
$i$ is further projected on the planes spanned by the 2 adjacent jets
$k$ and $j$. In order to be able to compare the different inter-jet
regions, the angle $\phi_i$ between the projected momentum and the
most energetic jet, is rescaled by the angle $\phi_{kj}$ between the
nearest jets. 

DELPHI and L3 perform a topological selection with tight cuts on the
angles of the jets in order to obtain a high purity sample with almost
planar jets, but with a low selection efficiency. ALEPH and OPAL use a
selection based on there standard W mass selection, which has a larger
efficiency but the jet topology is less trivial and the background
contamination is higher. 

\begin{floatingfigure}[r]{7.5cm}
  \centerline{\epsfig{file=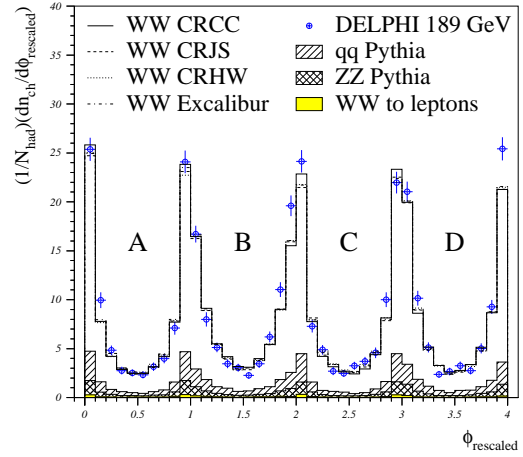,width=0.42\textwidth}}
\caption{The particle flow distribution as a function of the
  rescaled angle from DELPHI.}
\label{fig:flow}
\end{floatingfigure}

\begin{figure}[h!]
  \centerline{\epsfig{file=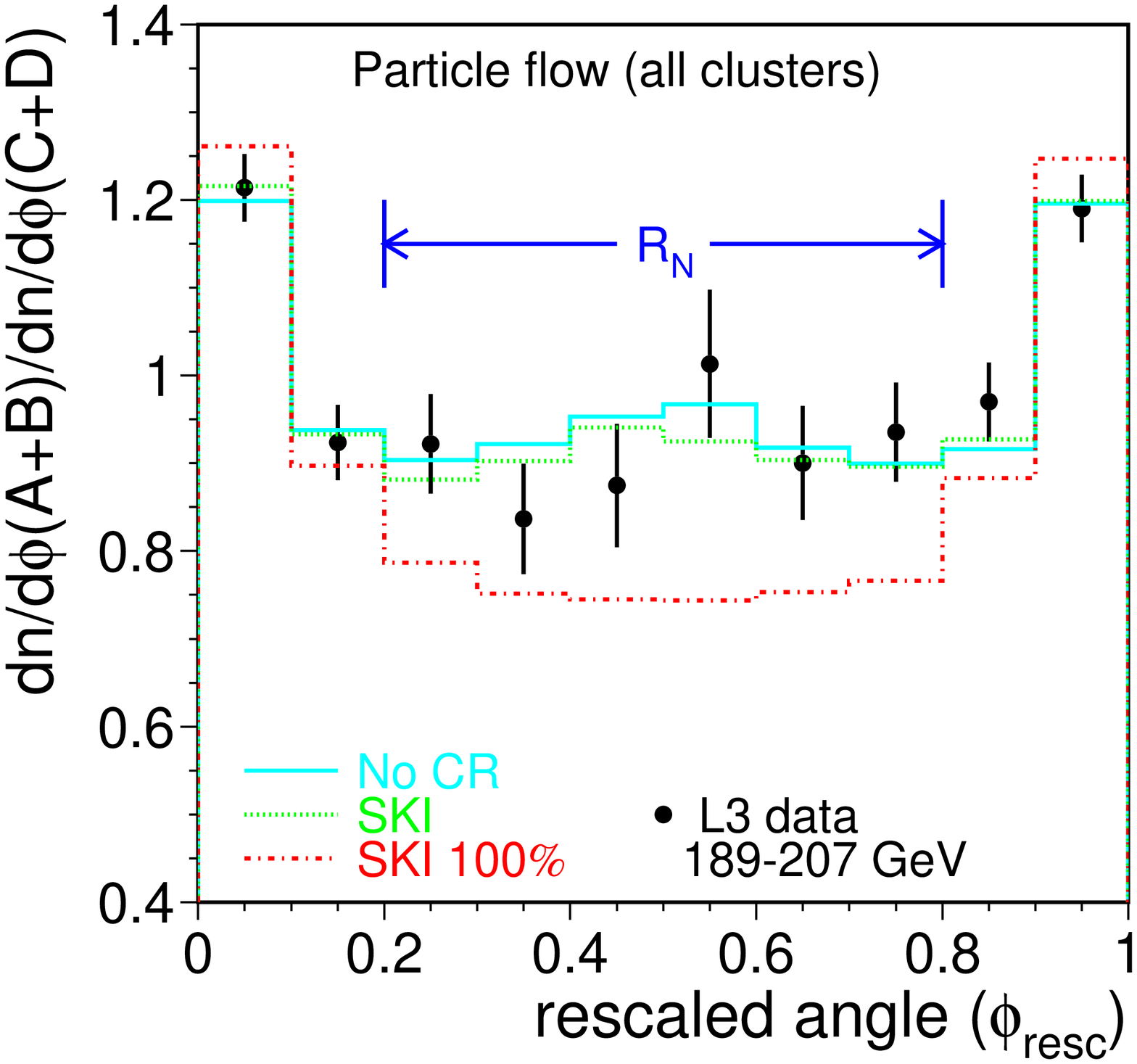,width=0.4\textwidth}
    \epsfig{file=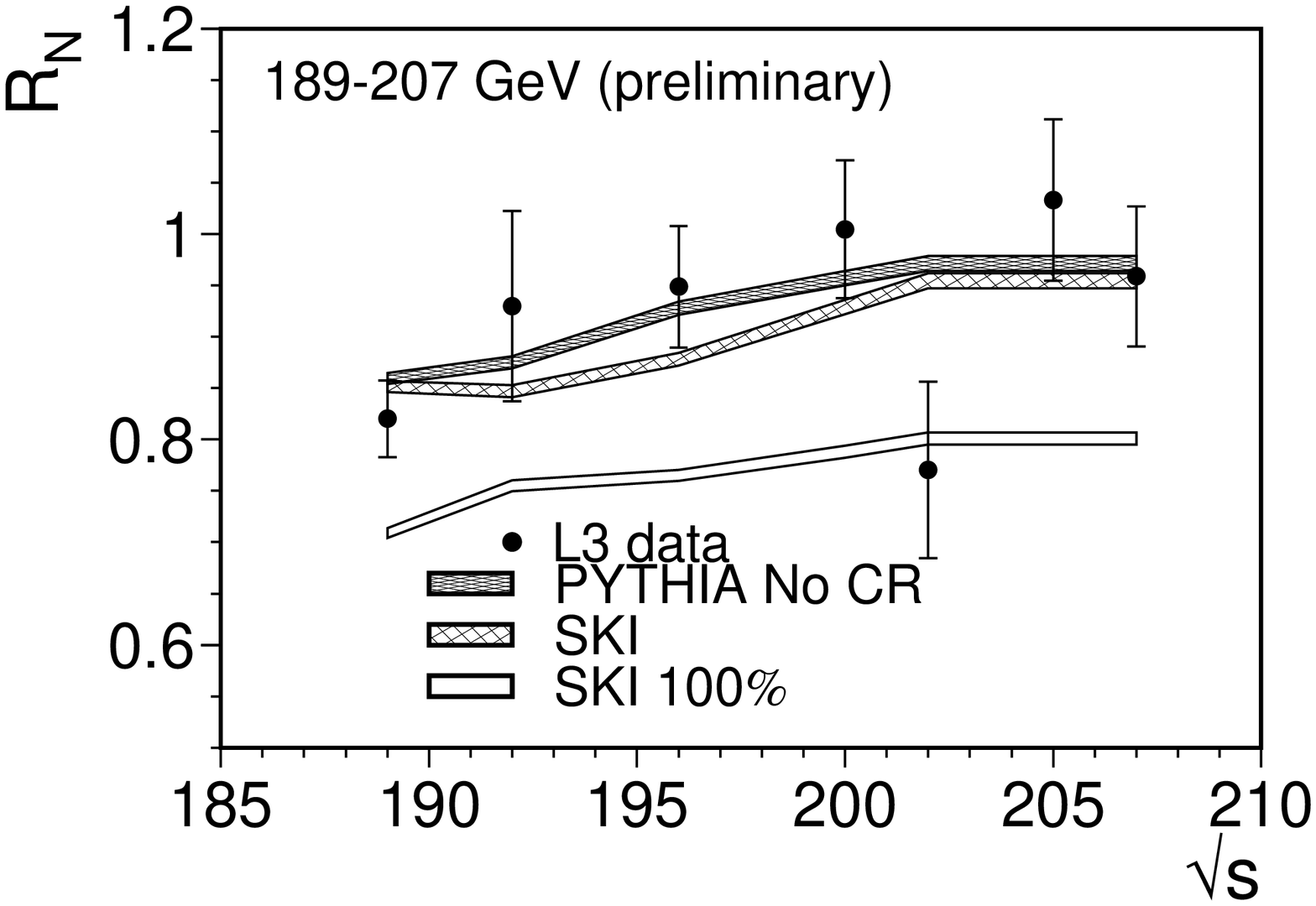,width=0.46\textwidth}}
\caption{The ratio $R$ between
  the particle flows inside and between the W's (left) 
  and the ratio $R_N$ between the integrals over the particle flows
  as function of centre-of-mass energy (right).}
\label{fig:ratio}
\end{figure}

An example of the particle flow distribution is shown in
figure \ref{fig:flow}. Due to the definition of the rescaled angle,
the direction of the jets are pointing to integer values. The two
regions inside a W (indicated with A and C on the plot) are now combined, 
as well as the two regions between the W's (B and D). These two 
particle flow distributions are then compared to each other by taking
the ratio, $R$, which can be seen in the left plot of figure
\ref{fig:ratio}. In order to quantify the CR effects of the 
different models, the particle flow inside and between the W's is
integrated over $\phi_{resc}$ between $0.2$ and $0.8$, the region most
sensitive to CR. Then the ratio between the two integrals is
called $R_N$ and shown in figure \ref{fig:ratio}.

In order to compare the different experiments, the ratios between the
$R_N$ value from data and a MC without and with CR (SKI with $100\%$
reconnection probability) have been  
calculated~\cite{abreu} and are shown in table \ref{tab:cr}. A
deviation from unity is a measure of the disagreement between data and
the model used. The
measurements of DELPHI and L3 show clearly a preference for no CR,
while ALEPH prefers a model inbetween no and $100\%$ CR. The OPAL
experiment comes to the same conclusion when using the W mass
selection. They also applied the topological selection as a cross
check analysis and then found a preference for no CR. This is actually
under study in all LEP 
experiments to see whether this is due to the type of selection, the
background or other possible sources.

The results can now be interpreted in terms of the free parameter
$k_I$ in the SKI model, see equation (\ref{eq:ski}), by calculating a
$\chi^2$ from the comparison of $R_N$ between data and MC samples
evaluated at different values of $k_I$. L3 obtains a limit of $k_I <
1.55$ at $68 \%$ C.L., while ALEPH prefers a value of $\langle k_I
\rangle = 3.55$ (a reconnection probability at $\sqrt{s} = 189$ GeV of
about $50\%$ and $80\%$ respectively).

\begin{table}[t]
  \caption{The ratios between the $R_N$ variables from data and a MC
    sample, one without CR and one with CR using the SKI model with
    $100\%$ reconnection probability. The values between brackets
    indicate the deviation from unity.}
  \label{tab:cr}
  \vspace{0.4cm}
  \begin{center}  
    \begin{tabular}{|c|cc|cc|}
      \hline
      &  \multicolumn{2}{|c}{$\langle R_R \rangle \quad$Data/no CR} &
      \multicolumn{2}{|c|}{$\langle R_R \rangle \quad$  Data/CR}  \\
      \hline
      ALEPH & $0.961 \pm 0.012 \pm 0.007$ & $(-2.8 \sigma)$ &
      $1.041 \pm 0.013 \pm 0.008$ & $(+2.7 \sigma)$ \\ 
      DELPHI & $1.009 \pm 0.030 \pm 0.019$ & $(+0.3 \sigma)$ & 
      $1.110 \pm 0.033 \pm 0.029$ & $(+2.5 \sigma)$ \\ 
      L3 & $0.990 \pm 0.025 \pm 0.023$ & $(-0.3 \sigma)$ &
      $1.194 \pm 0.039 \pm 0.029$ & $(+4.7 \sigma)$ \\
      OPAL & $0.906 \pm 0.033 \pm 0.011$ & $(-2.7 \sigma)$ &
      $1.050 \pm 0.038 \pm 0.013$ & $(+1.2 \sigma)$ \\
      & $0.996 \pm 0.051 \pm 0.011$ & $(-0.1 \sigma)$ & 
      $1.193 \pm 0.061 \pm 0.014$ & $(+3.1 \sigma)$ \\
      \hline
    \end{tabular}
  \end{center}
\end{table}

\section{Conclusions}\label{sec:concl}

A large part of the data samples collected by the four LEP experiments
has been analysed to study possible final state interactions in
hadronically decaying W pair events. 
None of the experiments observe any indication for the existence of
Bose-Einstein correlation between 
particles coming from different W bosons. The LUBOEI model used to
simulate such an effect is strongly disfavoured. 
The results from the different colour reconnection analyses are
inconsistent. DELPHI and L3 prefer no CR, while ALEPH prefers a large
value for the reconnection probability. OPAL gets different results
depending on the selection used. All experiments are presently
investigating the possible causes.
When the foreseen combination of the results from the four
collaborations will be performed, 
the contribution to the systematic error on the W mass from the
effects described here will be reduced significantly.

\end{document}